# Carrier Transport in PbS Nanocrystal Conducting Polymer Composites

Andrew Watt*, Troy Eichman, Halina Rubinsztein-Dunlop, and Paul Meredith

Soft Condensed Matter Physics Group and Centre for Biophotonics and Laser Science, School of Physical Sciences, University of Queensland, Brisbane, Australia. Fax: +61 7 3365 1242 Tel: +61 7 3365 1245;

*E-mail: watt@physics.uq.edu.au

PACS number(s): 73.63.Kv, 73.50.Gr, 73.61.Ph, 73.50.Pz.

## Abstract

In this paper we report the first measurements of carrier mobilities in an inorganic nanocrystal: conducting polymer composite. The composite material in question (lead sulphide nanocrystals in the conducting polymer poly (2-methoxy-5-(2'-ethyl-hexyloxy)-*p*-phenylene vinylene (MEH-PPV)) was made using a new single-pot, surfactant-free synthesis. Mobilties were measured using time of flight (ToF) and steady-state techniques. We have found that the inclusion of PbS nanocrystals in MEH-PPV both balances and markedly increases the hole and electron mobilities - the hole mobility is increased by a factor of $\sim 10^5$ and the electron mobility increased by $\sim 10^7$ under an applied bias of 5kVcm$^{-1}$. These results explain why dramatic improvements in electrical conductivity and photovoltaic performance are seen in devices fabricated from these composites.

## Introduction

Hole conducting polymers blended with an electron acceptor have been shown to be a promising system for solar energy conversion. The most commonly used electron acceptors are C60 derivatives[1] and cadmium selenide nanocrystals.[2]

Recently we developed a new approach for growing lead sulphide (PbS) nanocrystals directly in a conducting polymer to act as a charge acceptor.[3] The advantage of our new synthesis is that it eliminates the need for an initial insulating surfactant required to template the nanocrystal growth. In conventional methods this surfactant has to be removed before mixing with the conducting polymer since it inhibits efficient charge separation and transfer between the two components.[4] This removal can result in nanocrystal aggregation leading to poor material interfacing and therefore poor device efficiency.

For a donor-acceptor system to be a feasible photovoltaic material there are two important requirements: Firstly, electronic transport efficiency must be increased by balancing carrier distribution and minimizing loss through charge trapping and recombination. This can be achieved by matching the donor acceptor components[5] and creating electronically ordered materials with good donor acceptor interfaces.[6] Secondly, optical absorption must encompass as much of the solar spectrum as possible. This can be achieved by tuning the absorption properties of donor acceptor material[7] or by adding a third absorber.[8]

Lead sulphide nanocrystals were chosen because they offer improvements in several of these key aspects. Firstly, there is electronic enhancement of the system since PbS has equally mobile electrons and holes[9], unlike C60 and cadmium selenide where electron transport dominates. Secondly, the electron affinity ($\chi$) of bulk PbS is $\chi=3.3$eV which is larger than C60 ($\chi=2.6$eV), this increases the probability of charge separation.[10] Thirdly, the spectral absorption coverage is also improved when using PbS nanocrystals, because in the quantum regime (i.e. particle size smaller than the Bohr radius) PbS has tunable broad band absorption.[11] Finally, the excited state lifetime has been shown to be long in lead sulphide nanocrystals,[12] increasing the probability of photoexcited excitons being swept away by a devices inbuilt electric field. This combination of properties make PbS nanocrystals a flexible charge acceptor and optically active component when blended with conducting polymers.

Recently, we have demonstrated viable photovoltaic devices based upon a PbS nanocrystal: MEH-PPV composite.[13] These devices have significantly enhanced power conversion efficiencies when compared to MEH-PPV only devices. In this current article we examine the fundamentals of charge transport using time of flight (ToF) and steady state current-voltage techniques. In so doing, we seek to explain these results, and further demsonstrate the potential of nanocrystal: conjugated polymer composites as optoelectronic materials. Although mobility measurements have been reported on inorganic nanocrystals[14] imbedded in an insulating polymer, we believe this to be the first report of measurements on inorganic semiconductor nanocrytsal conducting polymer bulk heterojunction materials.

**Experimental**

**1. Sample Preparation**

The nanocrystal: conducting polymer blend was prepared as previously described,[3] and a typical transmission electron microscopy image of the composite material is shown in Figure 1. This image confirms that the nanocrystals are non-aggregated with a typical average size of 4 ± 2nm. Thin films of the material were solvent cast onto oxygen plasma-treated indium-tin oxide (ITO) substrates, and left to dry overnight under vacuum. Film thicknesses were measured using a Tencor Alpha-Step 500 Surface Profilometer and found to be between 0.7µm and 1.5µm. These films were thicker than those typically used in photovoltaic devices in order to minimize capacitive effects and make ToF measurements easier to temporally resolve. Aluminum cathodes were deposited by thermal evaporation through a shadow mask at a vacuum better than $10^{-5}$ mbar. The active device area was approximately 0.04cm$^2$. For current-voltage measurements the devices were prepared in a similar manner, but with thinner active layers (60-100nm) and an additional thin layer of PEDOT:PSS between the MEH-PPV or composite and the ITO.[13]

**2. Time of Flight (ToF) Measurements**

In the technique of ToF, a sheet of carriers is created near one of the contacts by a short laser pulse. Under the influence of an applied electric field, these carriers drift toward the counter electrode resulting in a transient current through the device. Charge transit times can then be measured and carrier type selected by changing the polarity and magnitude of the applied electric field. This method has been used extensively to examine transport in C60 derivative: conducting polymer blends.[15]

In our experiment, samples were illuminated through the ITO electrode using a Nd:YAG frequency-tripled laser (Surelite, 355 nm, 5 ns pulse length, 10 Hz repetition rate). Samples were measured under flowing argon in an electrically shielded box. Carrier type was selected by applying a bias using a Kiethley 2400 Source Measurement Unit. The current transient produced by the laser pulse in the presence of the applied electric field was recorded using a Agilent Infineon 54832B oscilloscope. Mobilities were then calculated according to[15]:

$$\mu = \frac{d}{Et_{tr}} \quad (1)$$

Where $d$ is the thickness of the film, $E$ is the applied field, and $t_{tr}$ the charge carrier transit time.

**3. Steady State Current Voltage (I-V) Measurements**

The I-V behaviour of the devices were determined using the Kiethley 2400 SMU, and once again, samples were measured under flowing argon in an electrically shielded box. Under conditions where the carrier mobility is independent of electric field, the steady state current is directly proportional to carrier mobility, and the current through a material may be written as:[16]

$$I(V) = (9/8)\varepsilon\varepsilon_0 \mu V^2 / d^3 \quad (2)$$

Where $\varepsilon$ is the material dielectric constant, $\mu$ the carrier mobility, $V$ the applied bias and $d$ is the sample thickness. Using this relation, an average value of mobility can be calculated, or, in combination with equation 1, a value for $\varepsilon$ can be extracted.

## Results and Discussion

### 1. Time of Flight Measurements

Figure 2 shows typical electron and hole photocurrent transients for the MEH-PPV and the composite devices. From initial examination it is clear that, as expected, hole transport dominates over electron transport in the MEH-PPV device (the magnitude of the hole current is significantly greater than the electron current). In contrast, in the composite material the magnitudes of the electron and hole currents are similar implying balance transport. Furthermore, the photocurrent transients of MEH-PPV show a dispersive behaviour[15] without visible inflection. In this case, the transit time is taken as the point of intersection of the linear asymptotes when plotted in a double logarithmic plot. The composite on the other hand is more complex with more than one decay process (possibly indicative of several conduction pathways). The photocurrent transients are characteristic of non-dispersive charge transport; the transit time is taken as the inflection point in the decay of the current.

Using equation 1, the hole and electron mobilities can be calculated for the curves shown in figure 2. Under an applied electric field of 4945 Vcm$^{-1}$ we found the hole and electron mobilities of the MEH-PPV films to be $2.83 \times 10^{-3} \pm 1.1 \times 10^{-4}$ cm$^2$V$^{-1}$s$^{-1}$ and $1.8 \times 10^{-5} \pm 1 \times 10^{-6}$ cm$^2$V$^{-1}$s$^{-1}$ respectively. Likewise, for the composite material, hole and electron mobilities were $95.8 \pm 0.4$ cm$^2$V$^{-1}$s$^{-1}$ and $83.9 \pm 0.3$ cm$^2$V$^{-1}$s$^{-1}$ respectively. These results clearly show that the inclusion of PbS nanocrystals into the MEH-PPV films both balances and markedly increases the hole and electron mobilities.

The field dependence of the mobility provides further information about the carrier transport mechanism. Figure 3 shows the field dependence of the hole mobility for MEH-PPV and composite devices. The electric field dependence of the MEH-PPV device (figure 3a) fits the the Poole-Frenkel relationship well;[17]

$$\ln \mu = s\sqrt{E} \qquad (3)$$

Where, once again, $E$ is the appied field, and $s$ is a constant of proportionality. Equation 3 only holds for dispersive transport, and Yu et al[18] showed that $s$ determines the degree of intermolecular interaction and is an indication of disorder. In the case of the composite material, it is clear that mobility is directly proportional to electric field (figure 3b). This confirms that the composite material is non-dispersive and a much more electronically ordered system. These results are supported by our microscopy data (presented in reference 3) which showed a degree of physical order in composite thin films which we believe is related to electronic order.

### 2. Steady State Current Voltage Measurements

Figure 4 shows the dark I-V characteristics for a device made from the composite material. This can be fitted to the Shockley equation (solid line) even at high electric fields - unlike other nanocrystal: conducting polymer composites which need to account for space-charge limited effects.[19] The ideality factor for our composite is $n=1.15 \pm 0.1$ which is approaching ideal (n=1), and significantly better than previoulsy published CdSe nanocrystal: conducting polymer composites[19]. This is further evidence that space-charge limited effects are small and that our material is electronically ordered. Using this result, and because the electric field and mobility are directly proportional (figure 3b), equation 2 can be used to analyze the data shown in figure 4 to extract a value for the bulk dielectric constant.

Literature values for the dielectric constant of MEH-PPV are accepted to be around 3[20] while for PbS nanocrystals the value varies from 900 to 3500 and is strongly dependent on nanocrystal size and packing density.[21] In an attempt to extract a realistic value for the bulk dielectric constant of our composite material we have calculated $\varepsilon$ as a function of mobility using equation 2. From this (figure 5) it can be seen that a dielectric constant of ~2.7 best fits the mobilities measured using time of flight, and is consistent with MEH-PPV being the absorber, and the PbS nanocrystals merely acting as charge acceptors. One could speculate that excitation at the nanocrystal bandedge would yield a significantly higher dielectric constant.

Finally, it is worthy of note that Bakueva *et al.* recently showed that a PbS nanocrystal: conducting polymer composite material displayed negative capacitive effects.[22] These effects could potentially alter the charge profile in the composite, making interpreatation of transport results in such systems more complex than we have outlined in this article.

**Conclusions**

In conclusion, the electron and hole mobilities of MEH-PPV and a PbS nanocrystal: MEH-PPV composite made using a new surfactant free synthesis have been studied using ToF and steady-state techniques. Most significantly, we have found that the nanocrystals act to balance and dramatically increase both electron and hole mobilities. Additionally, the mobilities are much less field dependent in the composite, indicating increased electronic order. From the steady-state I-V characteristics and assuming non-dispersive transport we have calculated a vale for the bulk dielectric constant of the nanocrystal: MEH-PPV composite and found it be similar to that of the native polymer. We believe that transport in the composite system is due to two conduction pathways in the material: a polymer-nanocrystal donor-acceptor pathway, and a purely nanocrystal percolation pathway arising from the equal electron and hole mobilities in PbS.

To understand the charge transport further in these systems, we plan to study the dependence on temperature and excitation wavelength. We believe that the material can also be optimized by further fine tuning of its properties, e.g. nanocrystal size and concentration. Overall, we believe this optoelectronic material has great potential. Our long term goal for PbS nanocrystal conducting polymer composites is to realize broad solar absorption via nanocrystal absorption tuning and carrier multiplication as recently described by Schaller and Klimov.[23]

**Acknowledgements**

The work was funded by the Australian Research Council. AARW thanks the University of Queensland for an International Postgraduate Research Scholarship. We would like to acknowledge the help of Chris McNeil at Newcastle University, Australia with film thickness measurements and Tim Mcintyre, Chris Vale and Andrew White at the University of Queensland for experimental assistance.

# FIGURES

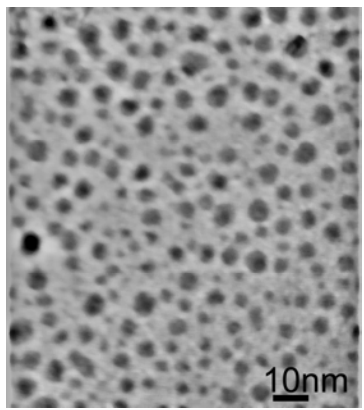

**FIG. 1.** Dark field transmission electron microscope image of an ensemble of nanocrystals average size 4 ± 2nm, prepared using our 1-pot surfactant-free synthesis.

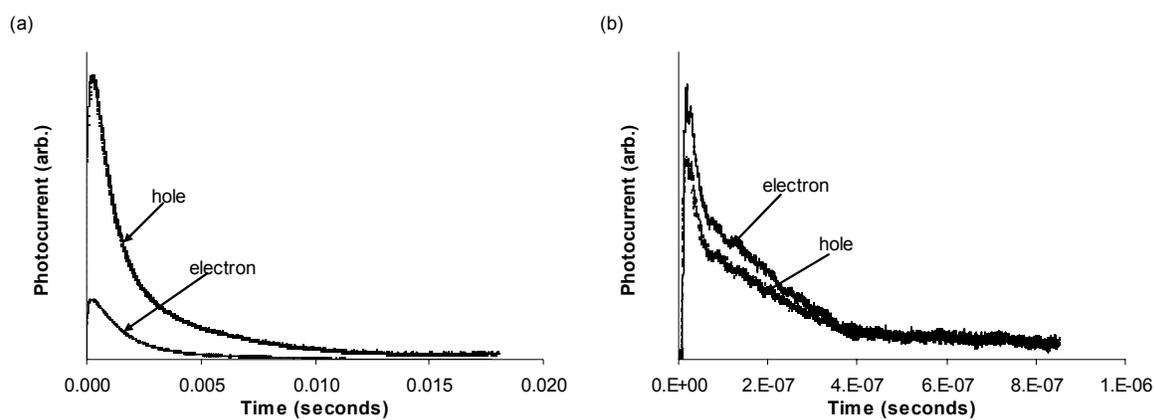

**FIG. 2.** Transient photocurent as a function of time for (a) a MEH-PPV device and (b) a PbS nanocrystal: MEH-PPV composite device.

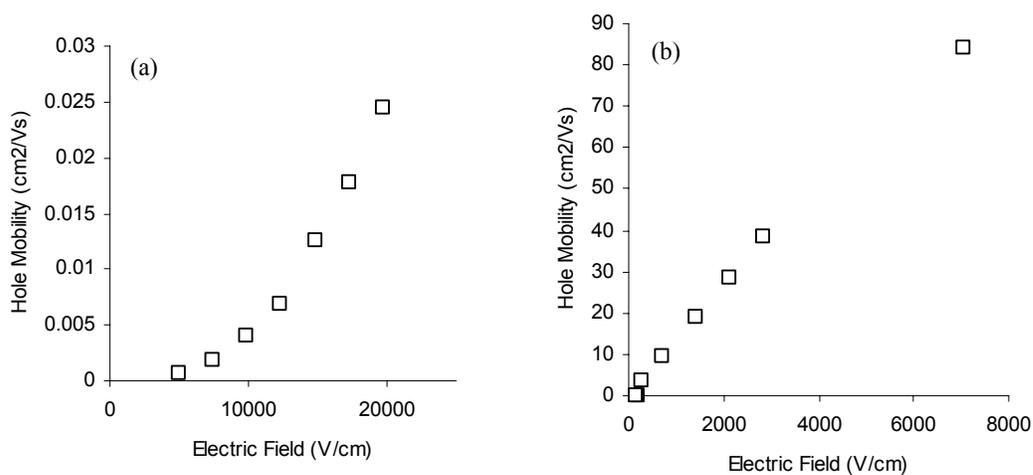

**FIG. 3** Experimental field dependence of hole mobility of (a) a MEH-PPV device and (b) a PbS nanocrystal: MEH-PPV composite device.

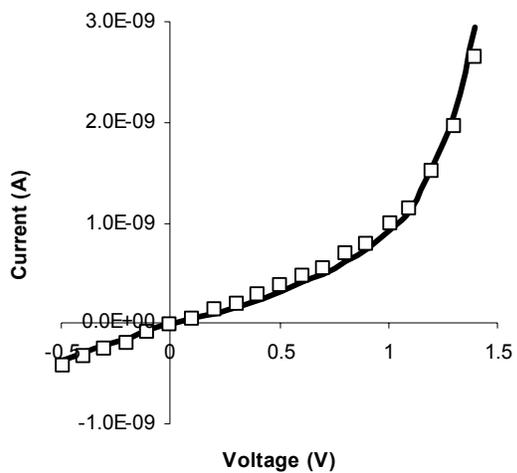

**FIG. 4.** Experimental dark current voltage (squares) and Schockley equation fit (line).

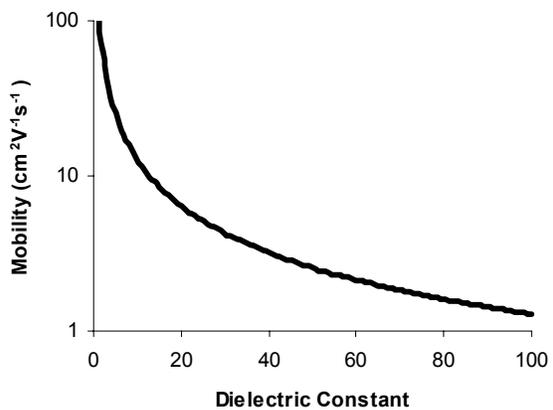

**FIG. 5.** Logarithmic-linear plots of mobility as a function of dielectric constant derived from steady state current voltage measurements.